\documentclass{aastex}
\usepackage{spr-astr-addons}
\usepackage{url}
\usepackage{graphicx}
\usepackage{bm}
\usepackage[misc]{ifsym}

\RequirePackage{color}

\begin{document}

\title{Review of Research on Lunar Dust Dynamics}
\slugcomment{Not to appear in Nonlearned J., 45.}
\shorttitle{Review on Lunar Dust Dynamics}
\shortauthors{Kun Yang et al.}

\author{Kun Yang\altaffilmark{1}} 
\and
\author{Weiming Feng\altaffilmark{1}$^{(\textrm{\Letter})}$}
\and 
\author{Luyuan Xu\altaffilmark{2,3}}
\and
\author{Xiaodong Liu\altaffilmark{4}$^{(\textrm{\Letter})}$}

\altaffiltext{1}{Department of Engineering Mechanics, Shandong University, 250061 Jinan, China}
\altaffiltext{2}{State Key Laboratory of Lunar and Planetary Sciences, Macau University of Science and Technology, Macau, China}
\altaffiltext{3}{CNSA Macau Center for Space Exploration and Science, Macau, China}
\altaffiltext{4}{School of Aeronautics and Astronautics, Sun Yat-sen University, Shenzhen Campus, 518107 Shenzhen, China}
\affil{Corresponding authors$^{(\textrm{\Letter})}$: \\ 
Weiming Feng (email: 
fwm@sdu.edu.cn)\\
Xiaodong Liu (email: liuxd36@mail.sysu.edu.cn) 
}

\begin{abstract}
Lunar dust particles are generated by hypervelocity impacts of interplanetary micron-meteoroids onto the surface of the Moon, which seriously threatens the security of explorations. Studying the lunar dust dynamics helps to understand the origin and migration mechanism of lunar dust, and to provide the theoretical guidelines for the orbital design of lunar space missions. This paper reviews previous research on the lunar dust dynamics, including the interplanetary impactor environment at the Earth-Moon system, the mass production rate, the initial mass, speed and ejecta angle distributions, the related space exploration missions, the dynamical model and spatial distribution of dust particles originating from the lunar surface in the whole Earth-Moon system.
\end{abstract}



\section{Introduction}
 
 Dust particles are widely distributed in the solar system. Giant planets, terrestrial planets, comets, and even some asteroids are surrounded by dust particles. Studies on distributions and dynamical behaviours of meteoroids (larger than $\sim30$ microns) and interplanetary dust particles (IDPs, smaller than $\sim30$ microns) in the solar system have been reviewed by \cite{koschny2019interplanetary}. Later \cite{janches2021meteoroids} summarized the research on the meteoroid bombardment process on airless bodies such as the Mercury and the Moon. The Earth-Moon system, with more exploration activities compared with other planets, is impacted by meteoroids and IDPs \citep{o2011review,grun2011lunar}. \cite{iglseder1996cosmic} proposed that there might be a dust cloud near the Moon, but the Munich dust counter (MDC) could not find the dust cloud around the Moon. A "horizon glow" near the lunar terminators was detected in both Surveyor and Apollo missions, which was considered to be lunar dust \citep{mccoy1976photometric,zook1991large}. The Lunar Dust Experiment (LDEX) carried by the Lunar Atmosphere and Dust Environment Explorer (LADEE) discovered a permanent and asymmetric dust cloud around the Moon \citep{horanyi2015permanent}. The density of this dust cloud is dependent on the local time, with a peak at $5$-$8$ hours local time. The typical size range of lunar particles is about from $0.3$ to $100\,\mathrm{\mu m}$, and follows a power-law distribution with the exponent $0.9$ \citep{horanyi2015permanent}, which is very close to the exponent $0.85$-$0.91$ obtained from results of the ground experiment \citep{buhl2014ejecta}. 
 
The Earth is surrounded by an atmosphere, so most of dust particles burn out before reaching the Earth's surface \citep{ceplecha1998meteor}. But the Moon is an airless body, which means it is directly exposed to the dusty environment \citep{hodges1975formation,morgan1989production}. When the IDPs and meteoroids impact the Moon at a high speed, dust particles are ejected from the lunar surface. Lunar dust particles migrate easily and affect the environment of the lunar surface \citep{colwell2007lunar} and near-lunar space \citep{grun2011lunar}. After long time orbital evolution, a fraction of the dust particles ejected from the lunar surface are widely distributed in the whole Earth-Moon system \citep{yang2021}. \cite{sliz2019celestial} found a dust cloud at the L5 point of the Earth-Moon system. In addition, the Space Dust (SPADUS) instrument on the Advanced Research and Global Observation Satellite (ARGOS) detected multiple dust impacts on the Earth orbit at a height of $850\,\mathrm{km}$ \citep{tuzzolino2001space,tuzzolino2005final}.

Lunar dust particles adversely affect the mechanical structure of spacecraft and even the health of astronauts \citep{turci2015free}. In Apollo missions, the lunar probe was attached by the lunar dust, resulting in the structural mechanical wear, seal failure, heat dissipation decline and other problems \citep{khan2008Lunar, cain2010Lunar}. \cite{sun2019mechanisms} reported the inflammatory reaction, fibrosis of blood vessels and bronchus induced by the lunar dust. 
 
 Studying the dynamics of the lunar dust helps to understand the space environment of the Earth-Moon system. In the orbital design stage, the dust flux impacting the spacecraft can be calculated in advance to evaluate the hazard of dust, which is of great help to the mission design. This paper reviews the research on the lunar dust, including the interplanetary impactor environment at the Earth-Moon system as well as the impactors' orbital characteristics, the mass yield, the initial dust distribution, the dynamic model of lunar dust and their spatial distribution characteristics in the Earth-Moon system.
 
 \section{The interplanetary impactor environment at the Earth-Moon system}

Dust particles originating from planetary moons in the solar system are produced by: (1) Impacts of IDPs and meteoroids on the surface of airless moons \citep{burns1984ethereal,kruger1999detection,janches2021meteoroids}; (2) Plume fountains of planetary moons \citep{stubbs2006dynamic}; (3) Space weathering and shedding of materials on the surface of moons, especially for airless ones \citep{clark2002asteroid}; (4) Electrostatic lofting for particles from the surfaces of moons without ionosphere \citep{criswell1973horizon}; (5) Volcanic ashes from moons \citep{graps2000io}. The origin of dust around each body differs, so do their generation mechanism and dynamical models. For the Moon, a significant number of dust particles are generated by the impacts of hypervelocity IDPs and interplanetary meteoroids onto the surface \citep{grun2011lunar}. Specifically, the interplanetary impactors can be divided into meteor showers and sporadic meteoroids while the flux of sporadic meteoroids is much greater compared with that of meteor showers \citep{Szalay2015Annual}. 

\begin{figure*}[t]
    \centering
    \includegraphics[width=0.8\textwidth]{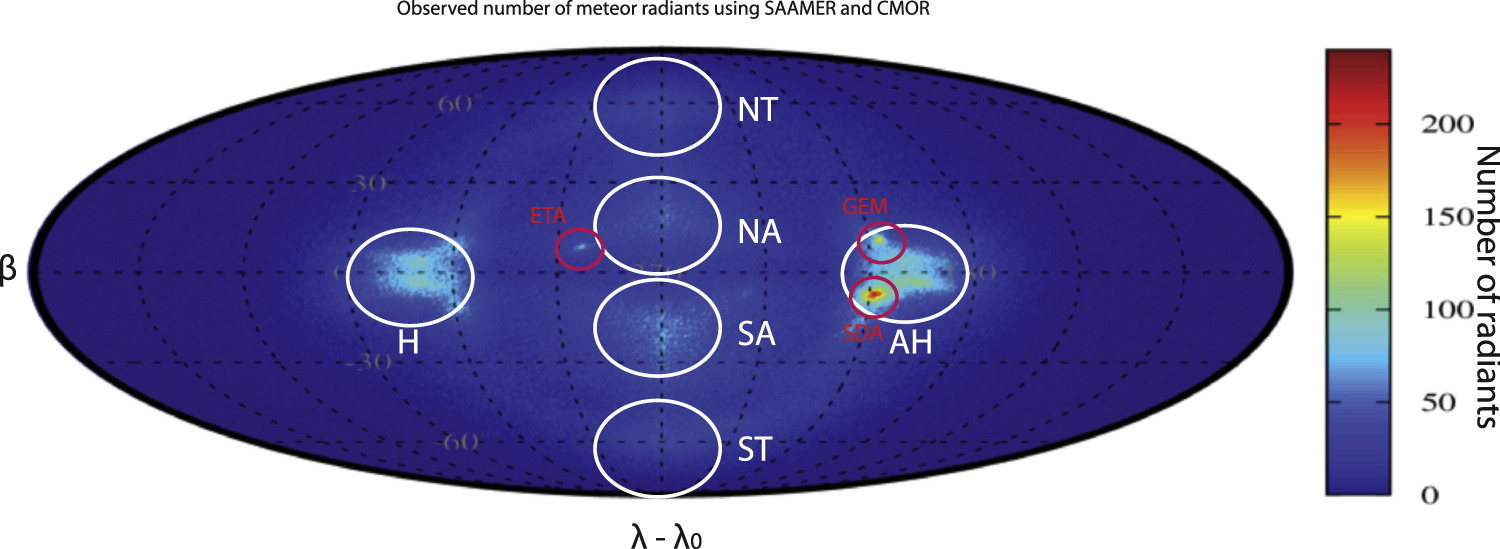}
    \caption{Sporadic meteor radiant distributions in the heliocentric ecliptic coordinate system, observed by the CMOR and the SAAMER. The symbol $\lambda$ denotes the heliocentric ecliptic longitude, and $\lambda_0$ denotes the true heliocentric longitude of the Sun. The white circles show the positions of Helion (H), Anti-Helion (AH), North Apex (NA), South Apex (SA), North Toroidal (NT), and South Toroidal (SA) sources, respectively. The red circles show three different meteor showers, ETA, GEM, and SDA, respectively. Reproduced with permission from \cite{janches2020comparative}, \copyright 2020 The Author(s).}
    \label{fig:RadiantDistributionOnSurface}
\end{figure*}

Sporadic meteoroids can be divided into six different sources: Helion, Anti-Helion, Apex, Anti-Apex and Northern/Southern Toroidal meteoroids \citep{jones1993sporadic}. \cite{Vogan1957} observed the sporadic meteoroids at the northern hemisphere of the Earth and found the flux of these meteoroids peaked at dawn during the summer months. \cite{weiss1957distribution} drew the same conclusion of the eruption of meteoroids in summer when observing them at the southern hemisphere of the Earth. In addition, the daily position of each meteoroid source changes little, only a few degrees from the average position, and the annual change is roughly the same \citep{campbell2006annual}. Note that this little variation of position is only 
visible in the heliocentric ecliptic coordinates. Based on long-term observations by the Canadian Meteor Orbit Radar (CMOR), the sporadic meteoroids were identified from the meteor streams \citep{jones2005canadian,brown2008meteoroid}. 
The distributions of meteoroid velocity, radiant and orbital elements were observed by the Southern Argentina Agile MEteor Radar \citep[SAAMER;][]{janches2015southern}. The results are consistent with previous observations by the CMOR \citep{campbell2008high}. \cite{pokorny2017orbital} analyzed more than one million meteoroid orbits observed by SAAMER during 2012 and 2015, and obtained the raw radiant distribution of these orbits in sun-centered ecliptic coordinates (see Fig.~3 in \citealt{pokorny2017orbital}). Fig.~\ref{fig:RadiantDistributionOnSurface} represents the sporadic meteor radiant distributions from the CMOR and SAAMER observations.

\subsection{Helion and Anti-Helion sources}

The Helion (HE) and Anti-Helion (AH) sporadic meteoroids were first discovered by the Jodrell Bank Observatory \citep{hawkins1956radio} and the Adelaide Radar \citep{weiss1957distribution}. HE and AH sources are concentrated in the ecliptic plane, at an angle of $60^\circ$-$70^\circ$ from the apex of the motion of Earth, close to the helion and antihelion points, respectively \citep{campbell2006annual}. \cite{poole1997structure} focused on the annual variability of the HE and AH sporadic meteor sources and found that the HE sources peaked in June and decreased sharply in July while the AH sources were maximized in June as well but reached a sub-peak in October. From the observation data by the CMOR, \cite{campbell2006annual} found that the annual variation of the HE sources was largest compared with that of other sources.

The orbital distributions of the HE and AH sources are similar to that of short-period comets, i.e. $a\approx1\,\mathrm{AU},\,e>0.3,\,i<30^\circ$, with a maximum speed between $20\,\mathrm{km\,s^{-1}}$ and $30\,\mathrm{km\,s^{-1}}$ \citep{jones2001modelling}. The parent bodies of the HE and AH sources are considered to be the Jupiter Family Comets (JFCs) and new comets, especially the 2P/Encke \citep{nesvorny2011dynamics}. Furthermore, the activity of JFCs itself cannot provide enough input for observed meteoroids, and disruptions of JFCs and Oort Cloud Comets (OCCs) may also supply the zodiacal dust cloud \citep{nesvorny2011dynamical, rigley2022comet}.

\subsection{Apex and Anti-Apex sources}

The Apex (AP) sources are also located in the ecliptic plane, which is consistent with the direction of the Earth's motion \citep{grun1980dynamics}. The AP sources show less annual variation compared with the HE and AH sources. The flux of the AP sources observed by CMOR is about twice that of the HE or AH sources \citep{campbell2006annual}. The AP sources move along retrograde orbits with small eccentricities, which is much different from the orbital characteristic of asteroids and most short-period comets \citep{campbell2008high}. The speed of the AP sources is high, showing a variation from the maximum $v\approx70\,\mathrm{km\,s^{-1}}$ at the ecliptic to the minimum $v\approx45\,\mathrm{km\,s^{-1}}$ at the north as well as south points. The velocity distribution is symmetrical about the ecliptic plane, with an average speed $v\approx60\,\mathrm{km\,s^{-1}}$ \citep{campbell2006annual}. The AP sources can be further divided into the north AP sources and the south AP sources \citep{sekanina1976statistical}, both moving along the direction of the motion of the Earth and located in approximately $20^\circ$ north and south from the ecliptic plane, respectively \citep{campbell2008high}. The parent bodies of the AP are considered to be Kreutz group comets \citep{andreev1992distribution}. The OCCs may also be the origin of AP sources through comparing the impact speeds and orbital elements of the OCCs and AP observations \citep{nesvorny2011dynamics}. Besides, the Halley type comets (HTCs) may also contribute to the AP sources \citep{pokorny2014dynamical} .

The Anti-Apex (AA) sources discovered by \cite{herschel1911observation} are also located in the ecliptic plane but move in the opposite direction of the AP sources. Compared with that of other sources, the flux of the AA sources is smallest while the mean semi-major axis ($a\approx3.7\pm 3.4\,\mathrm{AU}$) and eccentricity ($e\approx0.5\pm 0.3$) are relatively large \citep{campbell2008high, janches2000doppler, jones1993sporadic}.

\subsection{Toroidal sources}

The latest discovered toroidal sources include the north toroidal (NT) sources and the south toroidal (ST) sources, and they are located about $60^\circ$ north and south from the ecliptic plane, respectively \citep{elford1964meteor, jones1993sporadic}. The NT sources consist of dust particles on prograde orbits with $a\approx1\,\mathrm{AU}$ and $i\approx70^\circ$, with eccentricities and speeds peaked at $e\approx0.2$ and $v\approx35\,\mathrm{km\,s^{-1}}$, respectively \citep{campbell2008high}. \cite{pokorny2014dynamical} tracked the orbits of particles originated from the HTCs for millions of years, and proposed the NT sources observed by the CMOR to be possibly caused by activities and collisional break-ups of HTCs. The ST sources are very weak based on the data from the Advanced Meteor Orbit Radar \citep[AMOR;][]{galligan2005radiant}. However, later \cite{campbell2008high} believed the AMOR did not record many ST sources because of a cutoff of sizes. From the observations by the SAAMER, the ST sources and their meteor showers are strong \citep{janches2015southern, pokorny2017orbital, bruzzone2020comparative}. Conventionally, the ST and NT sources are always assumed to be symmetrical about the ecliptic plane with the same average strength \citep{szalay2019impact}.

\section{Detection missions}

A "horizon glow" near the lunar terminator was detected by the camera onboard Surveyor 5, 6 and 7 \citep{rennilson1974surveyor}. This interesting phenomenon was also noticed by Apollo 15 and 17, and believed to be forward sunlight scattered by the dust particles with $5\,\mathrm{\mu m}$ radius \citep{colwell2007lunar,glenar2011reanalysis}. The Lunokhod-2 rover measured the light brightness on the lunar surface and estimated that the dust cloud which scattered the sunlight was about $260\,\mathrm {m}$ above the lunar surface \citep{severny1975measurements}. Observations by detection missions above and the Lunar Ejecta and Meteorites Experiment (LEAM) onboard Apollo 17 \citep{berg1976lunar} indicated the existence of the lunar dust particles but failed to capture these particles.

On February 15 of 1992, the HITEN spacecraft developed by the Institute of Space and Astronautical Science of Japan (ISAS) was launched into a large-eccentricity ecliptic lunar orbit, with perilune of $100$-$8000\,\mathrm{km}$ and apolune of $50000\,\mathrm{km}$. The weight of the MDC onboard the HITEN is $605\,\mathrm{g}$ with a sensor area of $10\,\mathrm{cm} \times10\,\mathrm{cm}$, for detecting the IDPs near the Moon and calculating the mass and velocity of IDPs \citep{iglseder1993cosmic}. Before the HITEN spacecraft deliberately crashed into the lunar surface on April 10 of 1993, the MDC monitored $150$ impacts of particles with masses between $10^{-16}\,\mathrm{g}$ and $10^{-7}\,\mathrm{g}$ and speeds between $1.8\,\mathrm{km\,s^{-1}}$ and $100\,\mathrm{km\,s^{-1}}$. The measurements also show that the particles from the AP sources are larger in size than other sources ($>10^{-10}\,\mathrm{g}$), and the particles from the HE sources are relatively small ($<10^{-14}\,\mathrm{g} $) \citep{iglseder1996cosmic}.

\begin{figure}[t]
    \centering
    \includegraphics[width=0.9\columnwidth]{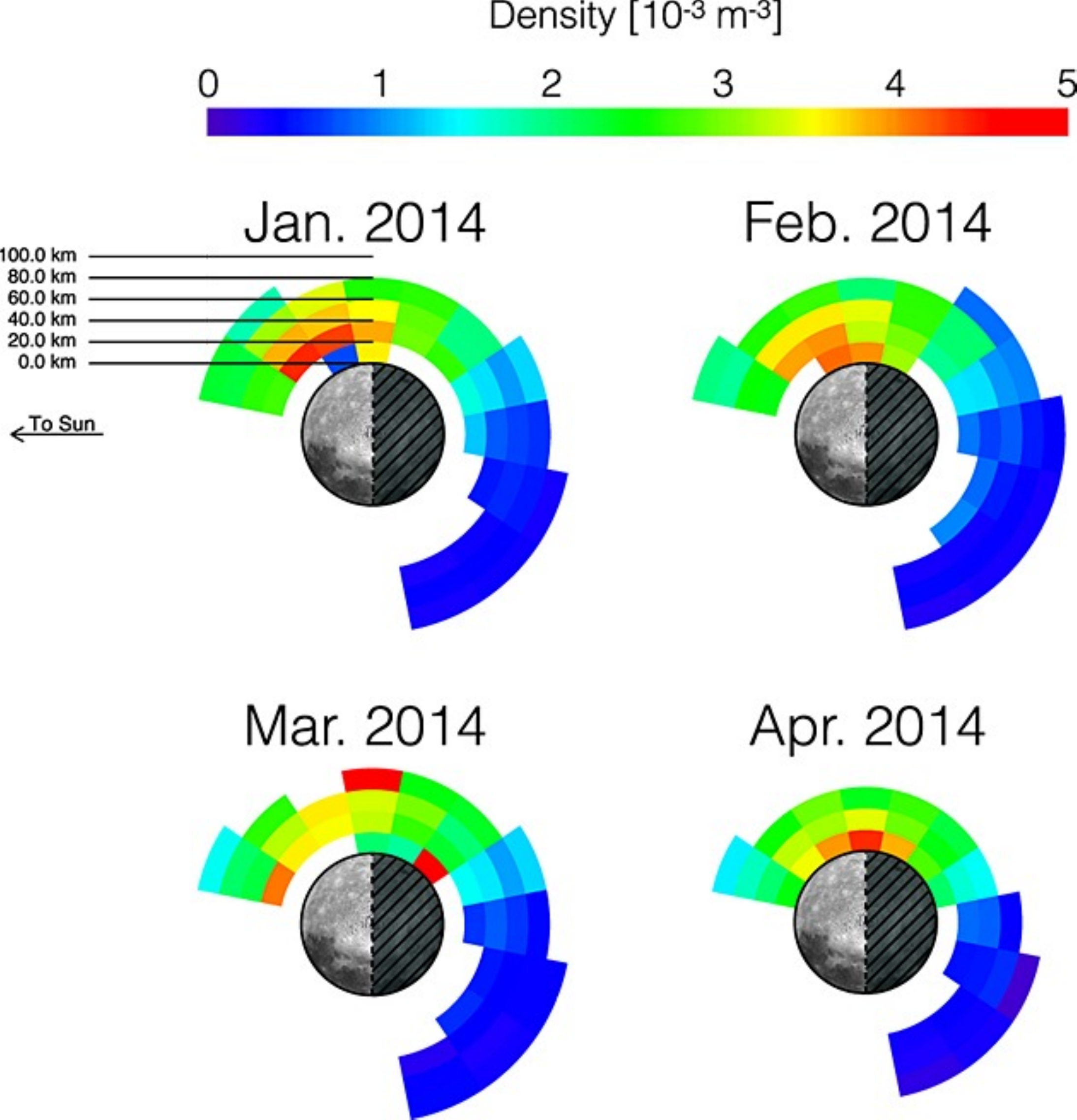}
    \caption{Average dust cloud density for different months in 2014 from LADEE data. Reproduced with permission from \cite{Szalay2015Annual}, \copyright 2015 American Geophysical Union.}
    \label{fig:clouddensity}
\end{figure}

However, due to the low sensitivity and the high orbit, the MDC did not detect the dust cloud near the Moon. The LDEX onboard the LADEE was launched on September 7 of 2013 and inserted into the lunar orbit after one month. The equipment inspection was carried out on the orbit with a period of approximately $40$ days at a distance $220$-$260\,\mathrm{km}$ from the Moon. After $150$ days, the LDEX was maneuvered into the lunar orbit with an altitude of $20$-$100\,\mathrm{km}$ and officially started to detect the lunar dust. During the whole mission period, the LDEX monitored about $14000$ impacts within $80$ days, about one impact outbreak per week ($10$-$50$ impacts per minute). Based on the LDEX data, \cite{horanyi2015permanent} reported a permanent and asymmetric secondary ejecta dust cloud, which tilted slightly towards the Sun, shown in Fig.~\ref{fig:clouddensity}. \cite{Szalay2015Annual} proposed that the ejecta dust cloud in the lunar equatorial plane was mainly generated by the impacts from the HE, AH and AP sources, and derived the different relative contributions for these three meteoroid sources. \cite{janches2018constraining} established a preliminary dynamical model for the lunar meteoroid environment to fit the LDEX measurements and argued that the HE and AH sources should have similar strengths. Later in an updated model by \citet{pokorny2019meteoroids}, the long-period comets (HTCs and OCCs) are found to be the main sources of the ejecta cloud around the Moon, and the JFCs plays a secondary important role. The meteoroid contributions in the model by \cite{pokorny2019meteoroids} cannot fully reproduce the configuration of the lunar dust cloud detected by the LDEX, and some possible reasons, such as the exposure to solar UV and solar wind \citep{janches2018constraining,pokorny2019meteoroids}, or additional sources \citep{szalay2020hyperbolic} may be necessary. However, this issue needs to be further investigated. The readers are also referred to Section 4 for more details on the analysis of the LDEX detections.

\begin{figure}[t]
    \centering
    \includegraphics[width=0.9\columnwidth]{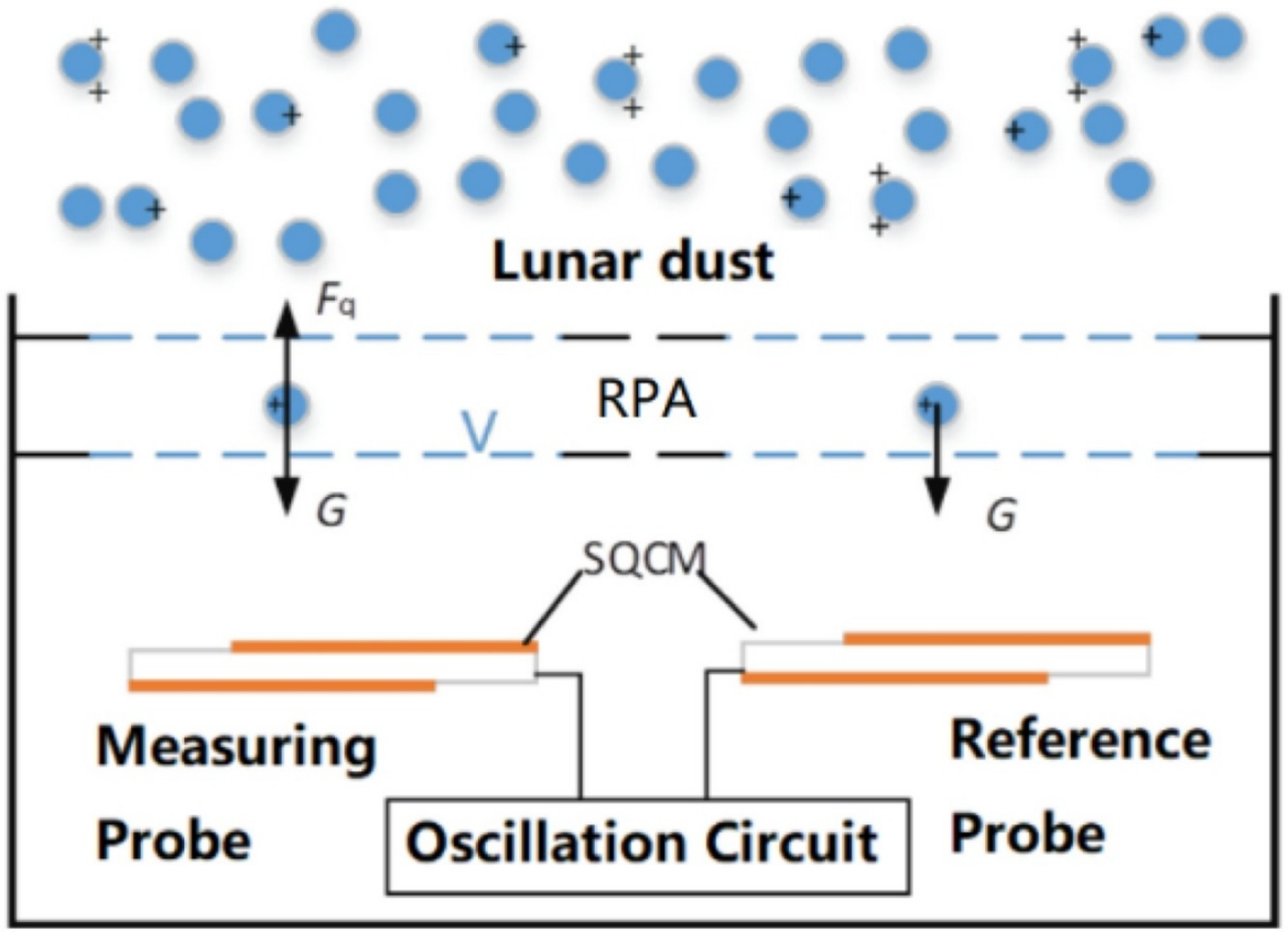}
    \caption{Design for the DCLD onboard Chang'e 5. Reproduced with permission from \cite{zhuang2021design}, \copyright 2021 Elsevier B.V.}
    \label{fig:DCLD}
\end{figure}

The lunar exploration project in China can be divided into three stages, 1) orbital exploration around the Moon; 2) soft-landing and deploying rovers; 3) sample-return. The Chang'e-3 lander arrived at the northern Mare Imbrium on December 14, 2013, equipped with a dust measuring instrument. The dust instrument is composed of two components: a Sticky Quartz Crystal Microbalance (SQCM) and a Solar Cell Probe \citep[SCP;][]{li2020situ}. After $364$ days, the detector obtained the amount of the lunar dust splashed during the landing period and the long-term accumulation of the lunar dust on the lunar surface. The amount of splashed dust within the first lunar day (mainly caused by the landing process) is $0.83\,\mathrm{mg\,cm^{-2}}$ \citep{zhang2020situ}, and the total accumulation mass induced by natural reasons is $0.0065\,\mathrm{mg\,cm^{-2}}$ \citep{li2019situ}, which is comparable to that from Apollo missions. The maximum active height of the dust is $28\,\mathrm{cm}$ in the Chang'e-3 landing area and the density decreases regularly with the increase of height, which is consistent with the estimation of dust electrostatic migration \citep{li2020situ}. By comparing with the activity area observed by the Apollo 17 mission, the dust activity range was found to not only vary with the lunar local time but also the geological characteristics of the landing area \citep{yan2019weak}. The Chang'e-5 lunar probe was launched on November 24 of 2020, carried a Detector of Charged Lunar Dust (DCLD), shown in Fig.~\ref{fig:DCLD}. Within 26 working hours, the charging characteristics and cumulative mass of the levitated lunar dust were measured \citep{zhuang2021design}.

More lunar exploration missions will be launched over the next decade. The Artemis I mission developed by NASA is scheduled to launch in June 2022, which is an uncrewed flight test for human lunar explorations\footnote{http://nasa.gov/artemis-1}. For the future Artemis series missions, the Orion spacecraft with crew will dock with a planned lunar space station, Lunar Orbital Platform - Gateway. The construction of Gateway will begin in 2023, which can provide abundant opportunities of long-term continuous measurements for lunar environment and future deep space explorations \citep{haws2019sls}. The Luna 25-27 missions are planned in Russia, the main tasks of which include exploring the lunar natural sources, estimating the lunar electromagnetic environment, studying the lunar surface topography, and analyzing the physical properties of the lunar regolith \citep{efanov2017moon}. The landers of Luna 25 and 27 will carry instruments to study the dusty plasma around the lunar surface \citep{popel2018lunar}. It is scheduled in India to launch the Chandrayaan 3 spacecraft in August 2022, which will be equipped with a lander and a rover for lunar polar explorations. Moreover, the Tanpopo experiment onboard the International Space Station aims to collect the interplanetary dust samples around the Low Earth Orbits (LEOs) for the later ground-based chemical analysis \citep{tabata2016ultralow}. The EQUilibriUm Lunar-Earth point 6U Spacecraft (EQUULEUS) will measure the impact flux of micrometeoroids over the cis-lunar region by an impact detector named Cis-Lunar Object detector within THermal insulation \citep[CLOTH;][]{funase2020mission}.

\section{Mass production rate and initial ejecta distribution}

Particles are generated from the surface of the Moon by the hypervelocity impacts of IDPs and meteoroids since the Moon is an airless celestial body \citep{pokorny2019meteoroids}. Every year approximately $10^6\,\mathrm{kg}$ IDPs hit the Moon, most of them range in $10\,\mathrm{nm}$-$1\,\mathrm{mm}$ with the speed $10$-$72\,\mathrm{km\,s^{-1}}$ and produce ejecta particles $1000$ times more than their own mass \citep{grun2011lunar}.

\begin{figure*}[t]
    \centering
    \includegraphics[width=0.8\textwidth]{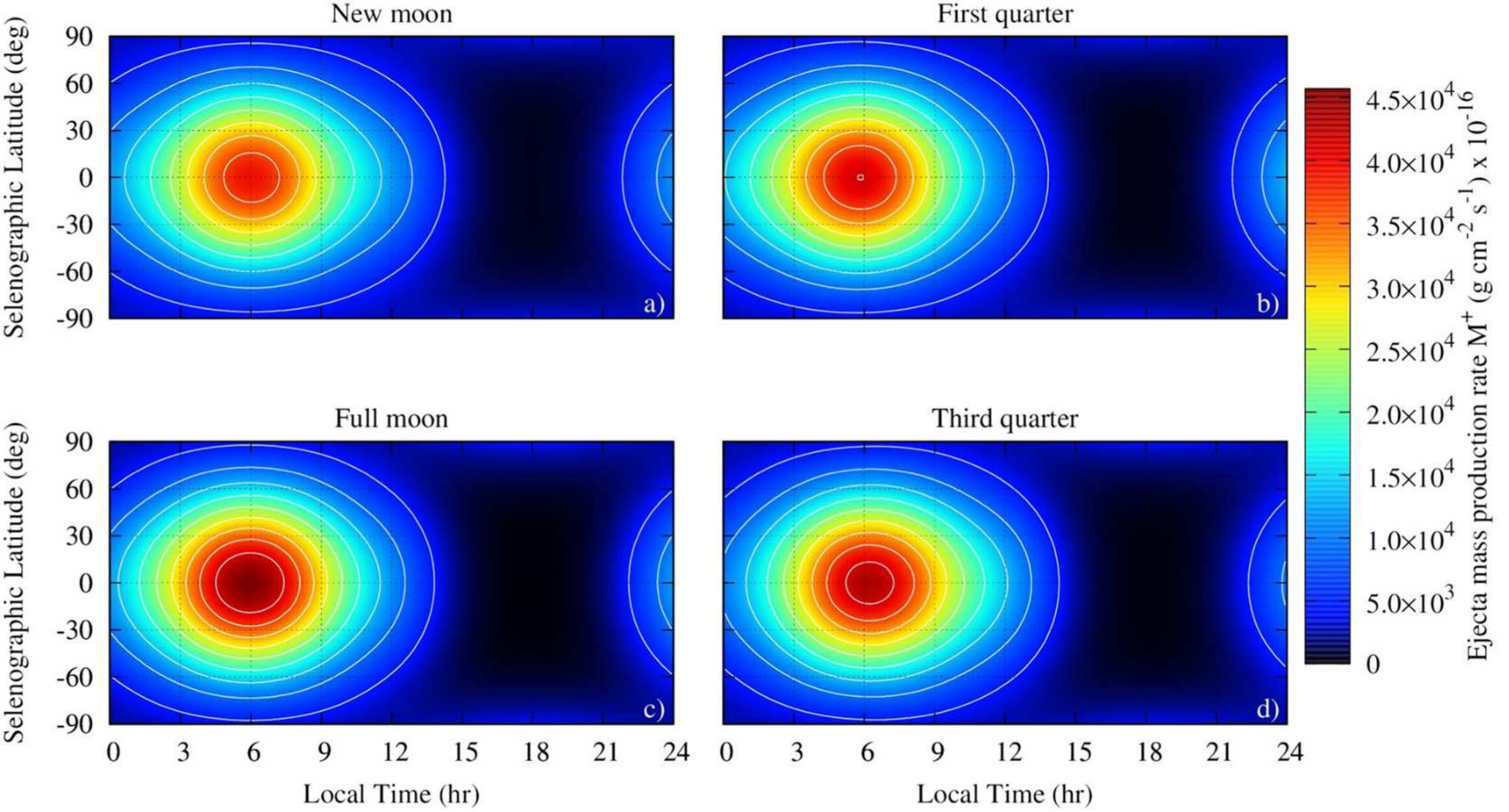}
    \caption{The mass production rate per unit surface $M^+$ for particles ejected from the lunar surface in the first lunation between July and August 2013. The contributions of individual sources, including main-belt asteroids (MBAs), JFCs, HTCs and OCCs, are obtained from LDEX measurements, and $M^+$ is calculated from Eq.~8 of \citet{pokorny2019meteoroids} (Eq.~\ref{M+latitude} in this review paper.) The horizontal axis denotes the local hours and the ordinate axis denotes the selenographic latitude in degrees. Reproduced with permission from \cite{pokorny2019meteoroids}, \copyright 2019 American Geophysical Union.}
    \label{fig:MassProductionRate}
\end{figure*}

Inferred from the experimental data, the ejecta yield $Y_p$, i.e. the mass ratio of the ejecta particles to the impactors, is a function of the material of the impacting target, the mass $m_\mathrm{imp}$, velocity $v_\mathrm{imp}$ and impacting angle $\phi$ of the impactors \citep{koschny2001impacts}
\begin{equation}
Y_\mathrm{p}\propto m_\mathrm{imp}^{\gamma_1} v_\mathrm{imp}^{\gamma_2}\cos^2\phi
\label{Yp}
\end{equation}
where $\gamma_1=0.23,\,\gamma_2=2.46$ from \cite{koschny2001impacts}. Based on the formula $M^+=F_\mathrm{imp}m_\mathrm{imp}Y_p$ \citep{krivov2003impact}, where $M^+$ is the mass production rate per unit surface and $F_\mathrm{imp}$ is the number flux of impactors, the Moon was assumed to be impacted by three meteoroid sources in the ecliptic plane (HE, AH and AP), and the mass production rate per unit surface for particles ejected from the Moon reads \citep{Szalay2015Annual}
\begin{small}
\begin{equation}
M^\mathrm{+}=C\sum_s\underbrace{F_{s}m_{s}^{\gamma_1+1}v_{s}^{\gamma_2}}_{w_s} \cos^3\left(\varphi-\varphi_s\right)\mathrm{H}\left(\pi/2-\left|\varphi-\varphi_s\right|\right) 
\label{threesources}
\end{equation}
\end{small}
where $C$ is the normalization constant, $s$ denotes the different meteoroid sources, $F_s$, $m_s$ and $v_s$ denote the flux, mass and impact speed for the $s$ source, respectively. The symbol $\varphi$ is the angle from the direction of the motion of Earth-Moon system (defined as the longitude), $\varphi_s$ is the longitude for each source, and $\mathrm{H}$ denotes the Heaviside function since impacting sources can only affect its reaching hemisphere of the Moon. The relative contributions of these three meteoroid sources were initially derived as $w_\mathrm{HE} = 0.37,\, w_\mathrm{AP} = 0.45,\,w_\mathrm{AH} = 0.18$ \citep{Szalay2015Annual}. Later, the HE and AH sources were argued to have comparable strengths based on the LDEX-derived data \citep{janches2018constraining}. The updated version of Eq.~\ref{threesources} after introducing the north and south toroidal sources obtained by \cite{szalay2019impact} reads 
\begin{equation}
M^\mathrm{+}=C\sum_s\underbrace{F_{s}m_{s}^{\gamma_1+1}v_{s}^{\gamma_2}}_{w_s} \cos^3\left(\Delta \phi_s\right)\mathrm{H}\left(\pi/2-\left|\Delta \phi_s\right|\right) 
\label{M+latitude}
\end{equation}
where the impact angle from the surface normal for each source $\Delta \phi_s$ is expressed as
\begin{equation}
    \Delta \phi_s=\cos^{-1}\left[\sin\lambda\sin\lambda_s\cos(\varphi-\varphi_s)+\cos\lambda\cos\lambda_s\right]
    \label{deltaphi}
\end{equation}
Here, $\lambda$ is the selenographic latitude and $\lambda_s$ is the latitude for each source. The relative contributions are $w_\mathrm{HE} = w_\mathrm{AH} = 0.198,\,w_\mathrm{AP} = 0.303,\,w_\mathrm{AA} = 0.025,\,w_\mathrm{NT}=w_\mathrm{ST}=0.138$. Note the strengths of HE and AH sources here are equal to each other, which is consistent with the results from \cite{janches2018constraining}. By integrating over the lunar surface, the mass production rate $M_\mathrm{total}^+$ caused by hypervelocity impacts is $18$ tons per day, corresponding to $0.2\,\mathrm{kg\,s^{-1}}$ \citep{szalay2019impact}. To make a better fit for both the dayside and nightside measurements from the LDEX, \cite{pokorny2019meteoroids} modified the mass production rate formula as
\begin{equation}
\begin{aligned}
&\left|LT-12\right|\textgreater6:\, M^+=\sum^{m_\mathrm{imp}}{Cm_\mathrm{imp}^{\gamma_1+1}v_\mathrm{imp}^{\gamma_2}}\\
&\left|LT-12\right|\leq6:\, M^+=\sum^{m_\mathrm{imp}}{Cm_\mathrm{imp}^{\gamma_1+1}v_\mathrm{imp}^{\gamma_2}}\\
&\qquad\qquad\qquad\qquad\qquad\times\left[1+\Psi\left(1-\left|12-LT\right|/6\right)\right]
\label{M+poko}
\end{aligned}
\end{equation}
where LT denotes the lunar local time and $\Psi\in[0,1]$ denotes the fractional increase factor between $6\,\mathrm{LT}$ and $12\,\mathrm{LT}$. Some possible reasons (exposure to solar UV and solar wind, and increase of surface temperature) was proposed for the enhancement of mass production rate on the dayside \citep{janches2018constraining,pokorny2019meteoroids}. Fig.~\ref{fig:MassProductionRate} shows the mass production rate per unit surface $M^+$ inferred from the Eq.~\ref{M+poko} during the first lunation in July and August 2013 \citep{pokorny2019meteoroids}. From the figure, the peak is canted towards the Sun at about $6\,\mathrm{LT}$. This can be explained by the dominant contribution of AP sources, whose density also peaks at around $6-8\,\mathrm{LT}$ \citep{Szalay2015Annual}. Besides, the impact-ejecta from meteoroids on hyperbolic trajectories that hit the Moon’s sunward side may be another reason for this asymmetric dust cloud \citep{szalay2020hyperbolic}. Based on \cite{pokorny2019meteoroids}, the meteoroid flux, the total energy deposition, and the ejecta mass were quantified in different topographically lunar polar regions, and the surface slope and the ejecta production rate were found to be positively correlated \citep{pokorny2020meteoroid}.
The number of grains ejected from the lunar surface for each size follows a power-law distribution \citep{kolokolova2004physical}
\begin{equation}
    N(>m)\propto m^{-\alpha}
\end{equation}
where $m$ is the mass of ejecta particles and $\alpha$ is the slope of the power law. The symbol $\alpha$ was believed to range from $0.5$-$0.9$ through the analysis for particles ejected from Phobos and Deimos \citep{krivov1998ethereal}. From the measurements of Ganymede \citep{kruger2000dust}, Callisto and Europa \citep{kruger2004jovian}, the exponent $\alpha\approx0.8$. \cite{buhl2014ejecta} implemented a series of ground-based impact experiments and obtained $\alpha\approx0.85$-$0.91$. The latest data from the LDEX onboard the LADEE indicates $\alpha\approx0.9$ \citep{horanyi2015permanent}, which is close to the values from other moons and ground test results. 

\begin{figure}[t]
    \centering
    \includegraphics[width=0.9\columnwidth]{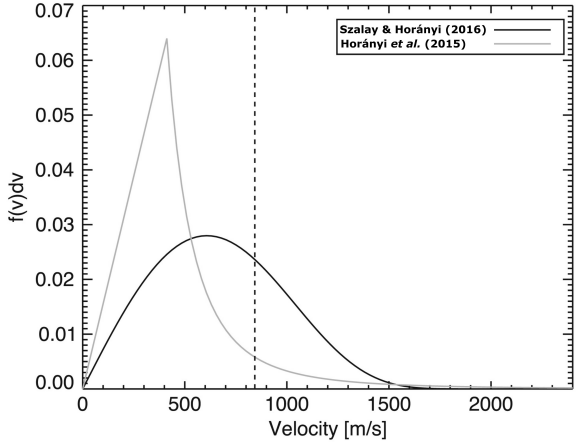}
    \caption{Comparison of the initial speed distribution model from \cite{Szalay2016Lunar} and \cite{horanyi2015permanent}. The speed distribution function of \cite{Szalay2016Lunar} is calculated from their Eq.~1 (Eq.~\ref{newspeeddistribution} in this review article), which differs from that of \cite{horanyi2015permanent} by removing the correlation between the local time and the altitude. Reproduced with permission from \cite{Szalay2016Lunar}, \copyright 2016 The Authors.}
    \label{fig:newspeeddistribution}
\end{figure}

By normalization in the speed range $[u_0,\infty]$, the initial speed distribution following a power law reads \citep{horanyi2015permanent}
\begin{equation}
    f_u(u)=\frac{\gamma}{u_0}\left(\frac{u}{u_0}\right)^{-\gamma-1}H(u-u_0)
\label{speeddistribution}
\end{equation}
Here, $u_0$ is the minimum ejecta speed, which can be estimated from the energy balance \citep{kruger2000dust}. The slope of power law $\gamma$ ranges within the interval $[2,3]$, determined by the materials of the target surface. For regolith-like soft targets like Phoebe, $\gamma=1.2$. For other harder surfaces like Enceladus and Rhea, $\gamma=2.0$ \citep{krivov2003impact}. Inferred from \cite{krivov2003impact}, a plausible slope of the power law for particles from the lunar surface is $\gamma=2.2$. Based on the LDEX measurements, \cite{Szalay2016Lunar} removed the local time dependence and obtained a new initial speed distribution for bound grains (Fig.~\ref{fig:newspeeddistribution})
\begin{equation}
    f_u(u)=\frac{2R_Mu}{h_0 u_e^2\left(1-\left(u/u_e\right)^2\right)^2}e^{-\frac{R_M/h_0}{\left(u_e/u\right)^2-1}}
\label{newspeeddistribution}
\end{equation}
where $R_M$ is the radius of the Moon, $u_e$ is the escape velocity of the Moon and $h_0=200\,\mathrm{km}$ is the scale height.

For the sizes of IDPs and meteoroids, the impact process is primarily governed by target properties (i.e., the strength-dominated regime). Based on extensive impact experiments and dimensional analyses, \cite{holsapple1993scaling} provided a comprehensive summary about the scaling of impact on solid planetary bodies. These scaling laws bridge the parameters between impactors (i.e., projectile diameter and density) and impact-related products (i.e., crater volume, depth, radius, and impact ejecta). For the scaling of ejecta, the review of \cite{holsapple1993scaling} briefly summarized the related contents in \cite{housen1983crater}, in which they described the scaling law of ejecta-related parameters such as ejection velocity and ejecta volume. However, since these estimates of the scaling relationship and parameters in \cite{housen1983crater} and later studies \citep[e.g.,][]{housen2011ejecta} are mainly based on the impact experiments of projectiles as small as several millimeters in diameter, the scaling of even smaller impactors such as IDPs may need to refer to further observations.

\begin{figure}[t]
    \centering
    \includegraphics[width=0.9\columnwidth]{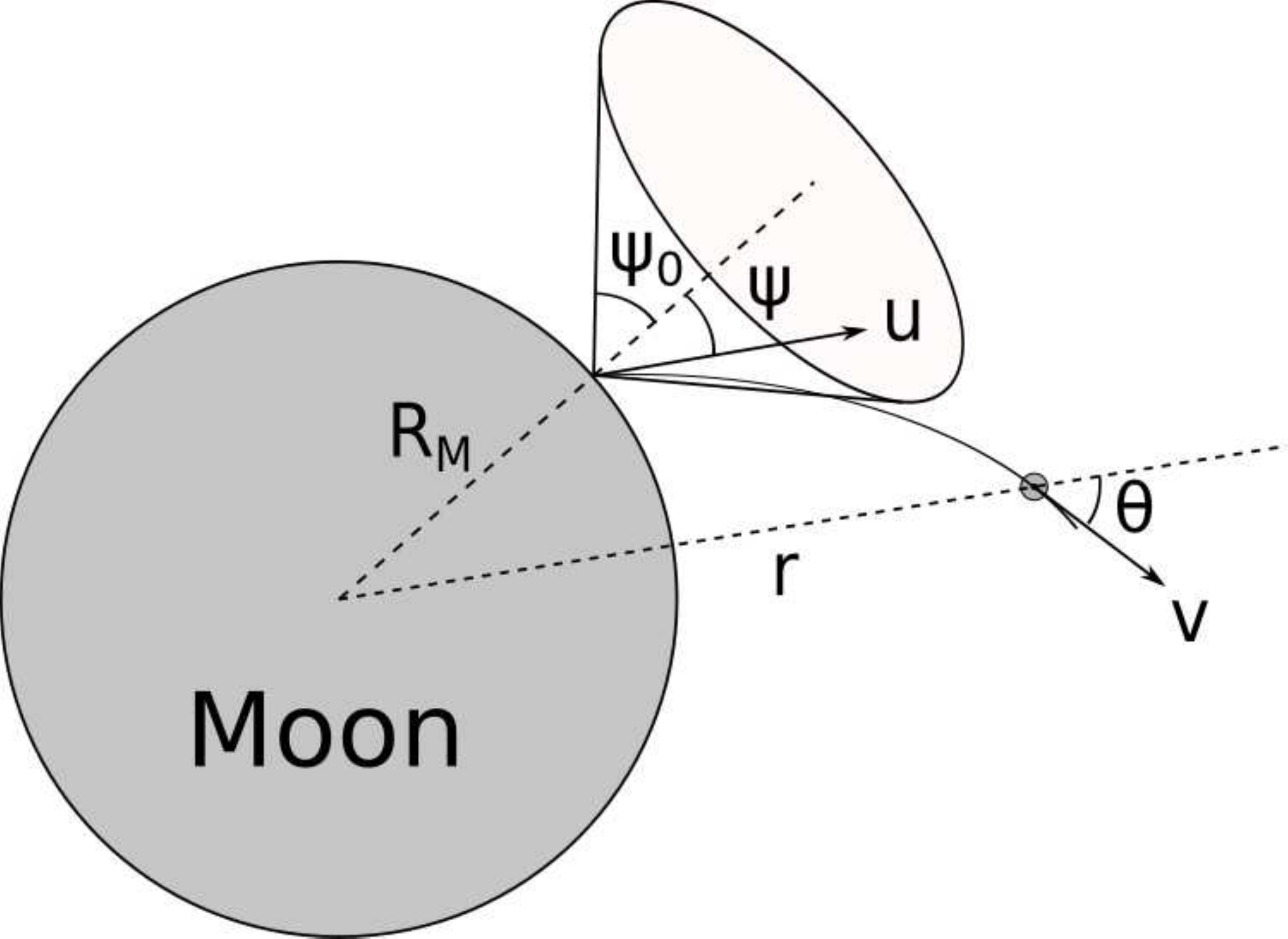}
    \caption{Impact-ejecta model. The symbol $R_M$ denotes the lunar radius and $\bm{r}$ denotes the position vector of the particle. The particle is ejected from the lunar surface with an initial speed $u$ and direction $\psi$ within a cone of half-opening angle $\psi_0\in[0,\pi/2]$.}
    \label{fig:EjetaModel}
\end{figure}

The initial ejecta angle $\psi$ is defined as the angle between the initial velocity vector and the normal of the lunar surface, and the ejecta particles are assumed to be uniformly distributed in a cone $\psi_0\in[0,\pi/2]$. By normalization to unity in the range $[0,\psi_0]$, the initial ejecta angle for grains ejected from the lunar surface follows \citep{horanyi2015permanent}
\begin{equation}
    f_\psi(\psi)=\frac{\sin{\psi}}{1-\cos{\psi_0}}H(\psi_0-\psi)
\label{angledistribution}
\end{equation}
The variables in Eq.~\ref{speeddistribution}, Eq.~\ref{newspeeddistribution} and Eq.~\ref{angledistribution} are shown in Fig.~\ref{fig:EjetaModel}. \cite{bernardoni2019impact} believed the dust cloud detected by LDEX was dominated by the reverse plume and the initial ejecta angle of the plume is
\begin{equation}
    f_\psi(\psi)=\frac{\sin\psi}{\cos \psi_1-\cos \psi_2}
\end{equation}
Here, the grains generated by hypervelocity impacts are supposed to ejecta in a narrow cone $[\psi_1,\psi_2]$. The opening angle for \cite{bernardoni2019impact} is $7^\circ8\pm3^\circ$, which is much narrower than the value ($30^\circ$) from \cite{horanyi2015permanent}.

The bombardment environments of micrometeoroids with $0.01$-$2\,\mathrm{mm}$ in diameter are distinct on the three major inner solar system airless bodies: Moon, Mercury, and Ceres. A comparative study of \cite{pokorny2021erosion} showed that Ceres suffers the mildest micrometeoroid bombardments with a total mass flux of $(4.5\pm1.2)\times10^{-17}\,\mathrm{kg\,m^{-2}\,s^{-1}}$, with $80\%$ more ejecta produced in the equatorial regions than in its polar regions. In comparison to Ceres, the Moon is bombarded by an average of $10$ times larger micrometeoroid mass flux and ejects $30$ times more ejecta, and Mercury suffers the most intense bombardments with $50$ times larger micrometeoroid mass flux and ejects $700$ times more ejecta on average.

Besides the primary impacts of projectiles with $10^{-3}\,\mathrm{\mu m}$ to $1\,\mathrm{km}$ in diameter, their secondary impactors (i.e., the fragments of ejecta produced from primary impact) also play an important role or even dominate in regolith rework and mixing (i.e., impact gardening) in the top few meters deep, which can effectively transport materials vertically and break the buried materials, e.g., water ice \citep{costello2018mixing,costello2020impact,costello2021secondary}. This impact gardening process can excavate, pulverize, and deplete the polar ice reserved in permanently shadowed regions (PSRs) near lunar and Mercury's poles \citep{costello2021secondary}, and the distinct impact gardening environments on Mercury and Moon may be partially responsible for the currently observed divergence between their ice-rich and ice-poor polar regions \citep{costello2020impact}.

\section{Dust distribution and dynamics}

The horizon-glow detected by Surveyor 7 was considered to be the forward sunlight scattered by charged particles ($5\,\mathrm{\mu m}$), which were levitated $3$-$30\,\mathrm{cm}$ above the lunar surface due to the electrostatic effect \citep{criswell1973horizon}. The strength of the electric field shows a substantial increase at the lunar terminator and the Earth's magnetic tail, so larger particles can be levitated in these areas \citep{colwell2007lunar}. \cite{hartzell2013dynamics} established a dynamical model to describe the motion of these levitated particles and pointed out that the initial speed of particles depended on the particle size and was insensitive to the mass of the central celestial body and the amount of charges carried by particles. \cite{xie2020Lunar} analyzed the LADEE measurements and found the dependence of the electrostatic migration height for levitated dust particles on the dust size. Particles with $0.1\,\mathrm{\mu m}$ radius can migrate to a height of tens of kilometers from the lunar surface while larger dust particles can only reach a few meters above the lunar surface due to less carried charges.

\begin{figure}[t]
    \centering
    \includegraphics[width=0.9\columnwidth]{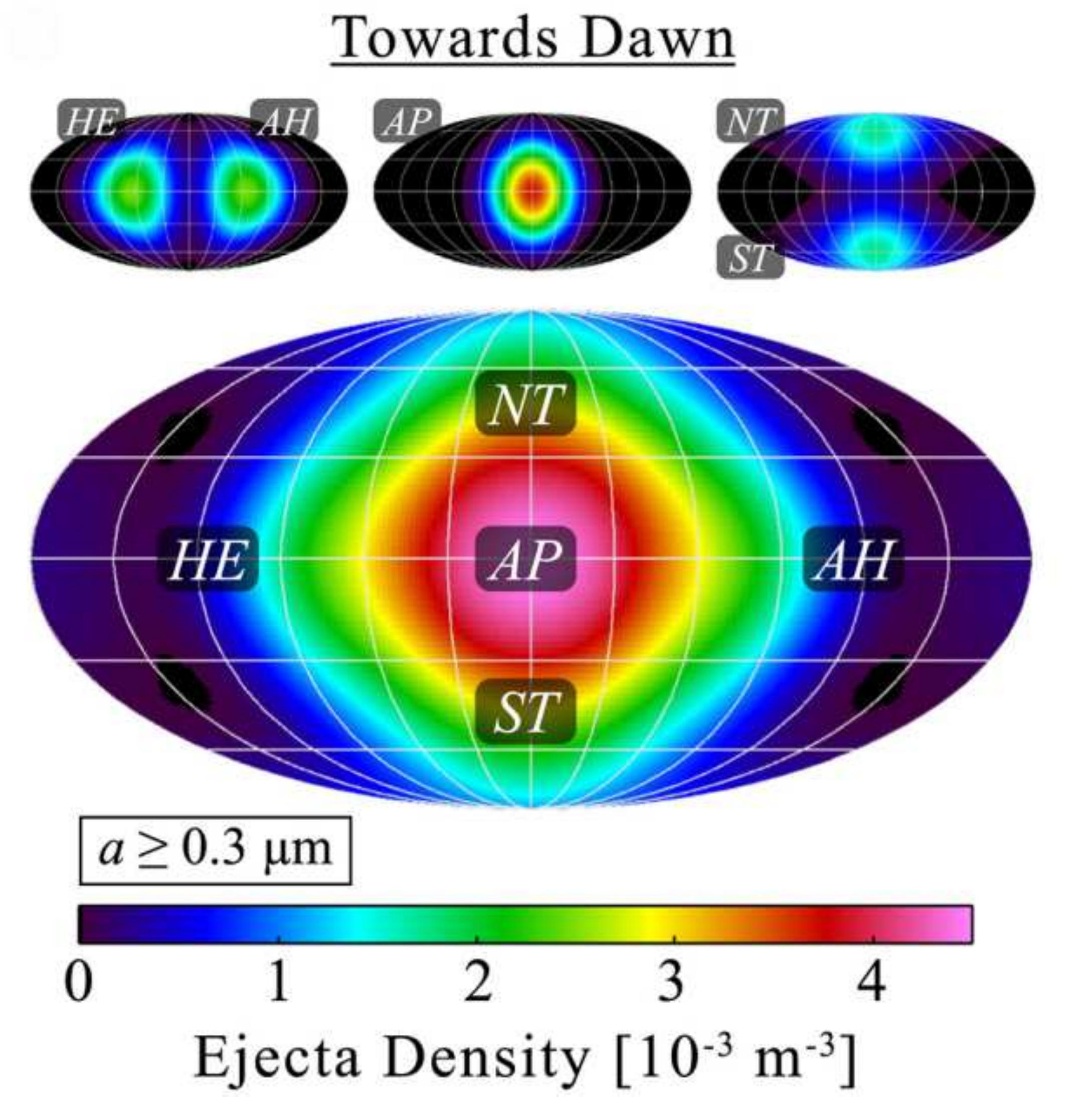}
    \caption{Ejecta number density at the lunar surface for each single source and the overall sources, calculated from Eq.~2 of \citet{szalay2019impact} (Eq.~\ref{numberdensity} in this review article, $h=0,a=0.3\,\mathrm{\mu m}$). Reproduced with permission from \cite{szalay2019impact}, \copyright 2018 American Geophysical Union.}
    \label{fig:EjetaDensity}
\end{figure}

From the LDEX data, the number density for lunar dust particles ($\textgreater0.3\,\mathrm{\mu m}$) is a function of altitude \citep{Szalay2016Lunar}
\begin{equation}
    n(h)=n_0 e^{-h/h_0}
\end{equation}
where $n_0$ is the number density at the lunar surface and $h$ is the altitude for particles. Considering the azimuthal, vertical and size dependence, the dust number density as a function of altitude $h$, longitude $\varphi$, latitude $\lambda$ and size $a$ reads \citep{szalay2019impact}
\begin{equation}
\begin{aligned}
n(h,\varphi,\lambda,a)=&e^{-h/h_0} a^{-q} n_w  \\
&\sum_s w_s \cos^3\left(\Delta \phi_s\right)\mathrm{H}\left(\pi/2-\left|\Delta \phi_s\right|\right) 
\end{aligned}
\label{numberdensity}
\end{equation}
where $q=2.7$ \citep{horanyi2015permanent}, $n_w$ is the normalized number density obtained by fitting with the LDEX measurements, and $\Delta \phi_s$ is a function of longitude $\varphi$ and latitude $\lambda$ for each source, expressed in Eq.~\ref{deltaphi}. Fig.~\ref{fig:EjetaDensity} shows the number density at the lunar surface.

 Most of dust particles fall back to the lunar surface \citep{Szalay2016Lunar}. A fraction of dust particles generated by hypervelocity impacts may escape from the lunar gravity and evolve into the Earth-Moon system. These particles are affected by various forces, including the gravity of Moon, Earth and Sun, the solar radiation pressure, the Poynting-Robertson drag and the Lorentz force, and the dynamical equation for lunar dust particles in the J2000 frame can be written as \citep[e.g.][]{murray1999solar, burns1979radiation, liu2021configuration}
\begin{equation}
\begin{aligned}
\ddot{\bm{r}} =&GM_E\nabla{\left\{\frac{1}{r}\left[1-J_2\left(\frac{R_E}{r}\right)^2P_2(\cos\theta)\right]\right\}}\\
+&GM_\mathrm{S}\left(\frac{\bm{r_{\mathrm{pS}}}}{r_{\mathrm{pS}}^3}-\frac{\bm{r_{\mathrm{S}}}}{r_{\mathrm{S}}^3}\right)+GM_\mathrm{M}\left(\frac{\bm{r_{\mathrm{pM}}}}{r_{\mathrm{pM}}^3}-\frac{\bm{r_{\mathrm{M}}}}{r_{\mathrm{M}}^3}\right) \\
+&\frac{3Q_\mathrm{c}Q_{\mathrm{pr}}\mathrm{AU}^2}{4\left(\bm{r}-\bm{r_\mathrm{S}}\right)^2\rho_\mathrm{g}r_\mathrm{g}c} \left[1-\frac{\left(\dot{\bm{r}}-\dot{\bm{r_\mathrm{S}}}\right)\cdot{\bm{\hat{r}_{\mathrm{Sp}}}}}{c}\right]\bm{\hat{r}_{\mathrm{Sp}}}\\
+&\frac{Q}{m_\mathrm{g}}\left(\bm{v_\mathrm{B}}\times \bm{B}\right)
\end{aligned} \label{eq:equations_of_motion}
\end{equation}
where $\bm{r}$ denotes the position vector of the dust particle, $G$ is the gravitational constant, $M_\mathrm{E}$ is the Earth mass, $R_\mathrm{E}$ is the Earth radius, $J_2$ is the second-degree zonal harmonic of the Earth's gravity field, $P_{2}$ is the second-degree Legendre function, $\theta$ is the colatitude in body-fixed frame, $M_\mathrm{S}$ denotes the Solar mass, $M_\mathrm{M}$ denotes the lunar mass, $\bm{r_{\mathrm{S}}}$ and $\bm{r_{\mathrm{M}}}$ denote the position vectors of Sun and Moon in the J$2000$ frame, $\bm{r_{\mathrm{pS}}}$ denotes the vector from the particle to the Sun, $\bm{r_{\mathrm{pM}}}$ denotes the vector from the particle to the Moon, $Q_\mathrm{c}$ is the solar radiation energy flux at $1\,\mathrm{AU}$ distance from the Sun, $Q_\mathrm{pr}$ is the radiation pressure efficiency factor, $\rho_\mathrm{g}$ is the grain density, $r_\mathrm{g}$ is the grain radius, $c$ denotes the light speed, $\hat{\bm{r_{\mathrm{Sp}}}}$ denotes the unit vector from the Sun to the particle, $Q$ is the grain charge, $m_\mathrm{g}$ is the grain mass, $\bm{v_\mathrm{B}}$ is the relative velocity of the grain to magnetic field and $\bm{B}$ is the strength of magnetic field.

The lunar gravity is significant since the particles are generated from the Moon. The second-degree zonal harmonic $J_2\approx1.082\times10^{-3}$ \citep{pavlis2012development} of the Earth's gravity field is important when the particles are close to the Earth. Besides, the solar gravitational perturbation is also non-negligible for particles in the Earth-Moon system. Meanwhile, dust particles can be blown out of the solar system due to the existence of solar photons, which is called the solar radiation pressure \citep{burns1979radiation}. The radiation pressure efficiency factor $Q_\mathrm{pr}$ depends on the size and material of dust particles, which can be calculated from Mie theory \citep{mishchenko1999bidirectional,mishchenko2002scattering}. The term related to the speed in the formula for solar radiation pressure is Poynting-Robertson drag, which can make particles spiral towards the Sun \citep{robertson1937dynamical}. Dust particles ejected from the lunar surface are exposed to the plasma and UV environment continuously, and carry electrostatic charges during the charging process \citep{horanyi1996charged}. The grain charge $Q\propto r_\mathrm{g}$ so the acceleration caused by the Lorentz force $\ddot{\bm{r}}_{\mathrm{L}}\propto 1/r_g^2$, i.e. the effect of Lorentz force plays a more important role for small dust particles. Particles originating from the lunar surface are moving in the interplanetary magnetic field (IMF), the Earth's magnetic field and the lunar magnetic field. The spiral model was considered a good approximation of the IMF \citep{parker1958dynamics}, and the average strength \citep{gustafson1994physics} and polarity was estimated \citep{landgraf2000modeling}. The strength of Earth's magnetic fields is usually detected by flyby or orbital spacecraft and the approximate average strength of the observed surface field for Earth is $5\times10^{-5}\,\mathrm{T}$ \citep{stevenson2003planetary}.

\begin{figure}
    \centering
    \includegraphics[width=0.9\columnwidth]{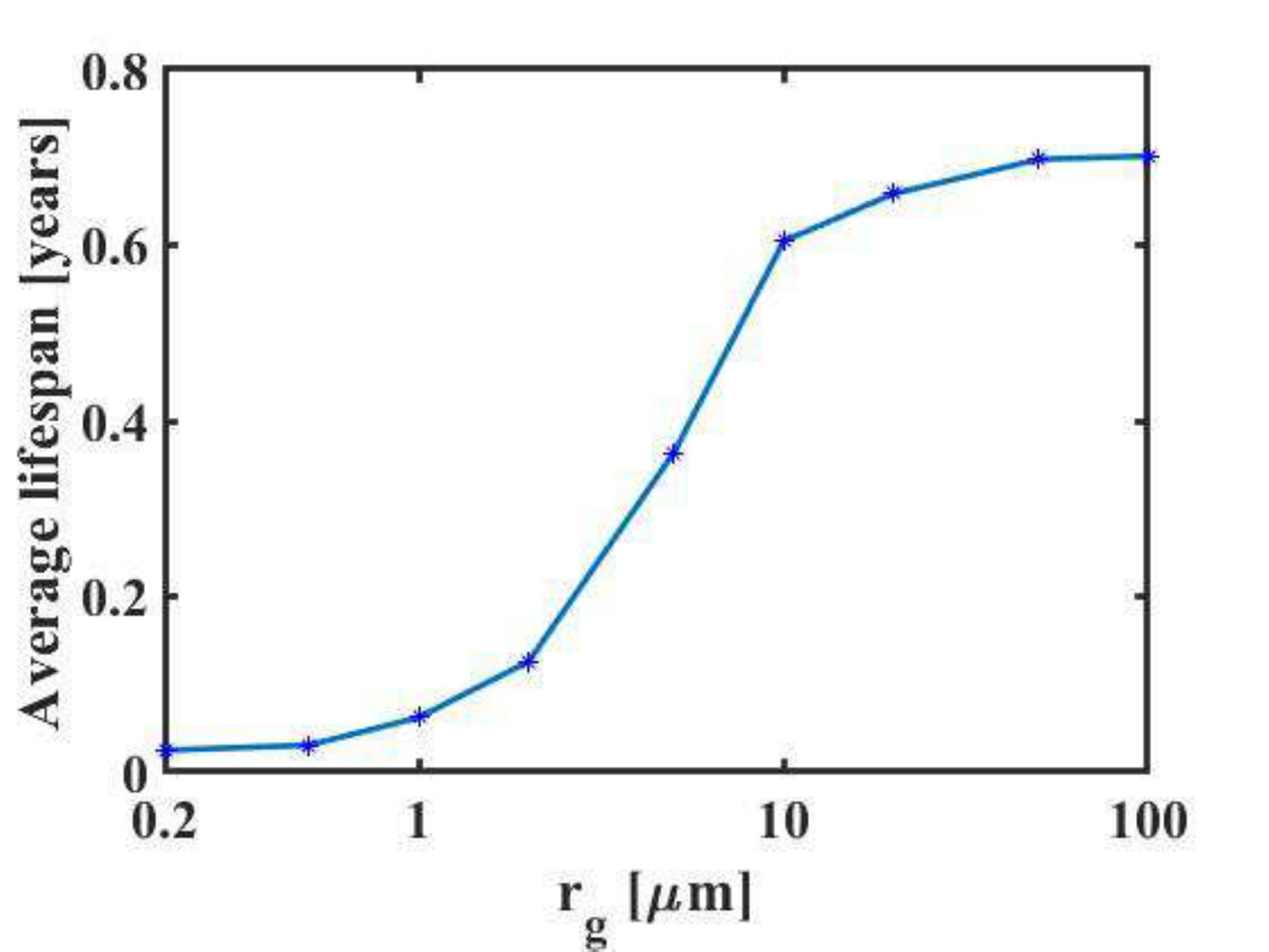}
    \caption{Average lifetimes for grains ejected from the lunar surface, obtained from numerical simulations. Reproduced with permission from \cite{yang2021}, \copyright ESO.}
    \label{fig:LifeSpan}
\end{figure}

\begin{figure}
    \centering
    \includegraphics[height=0.9\columnwidth,angle=90]{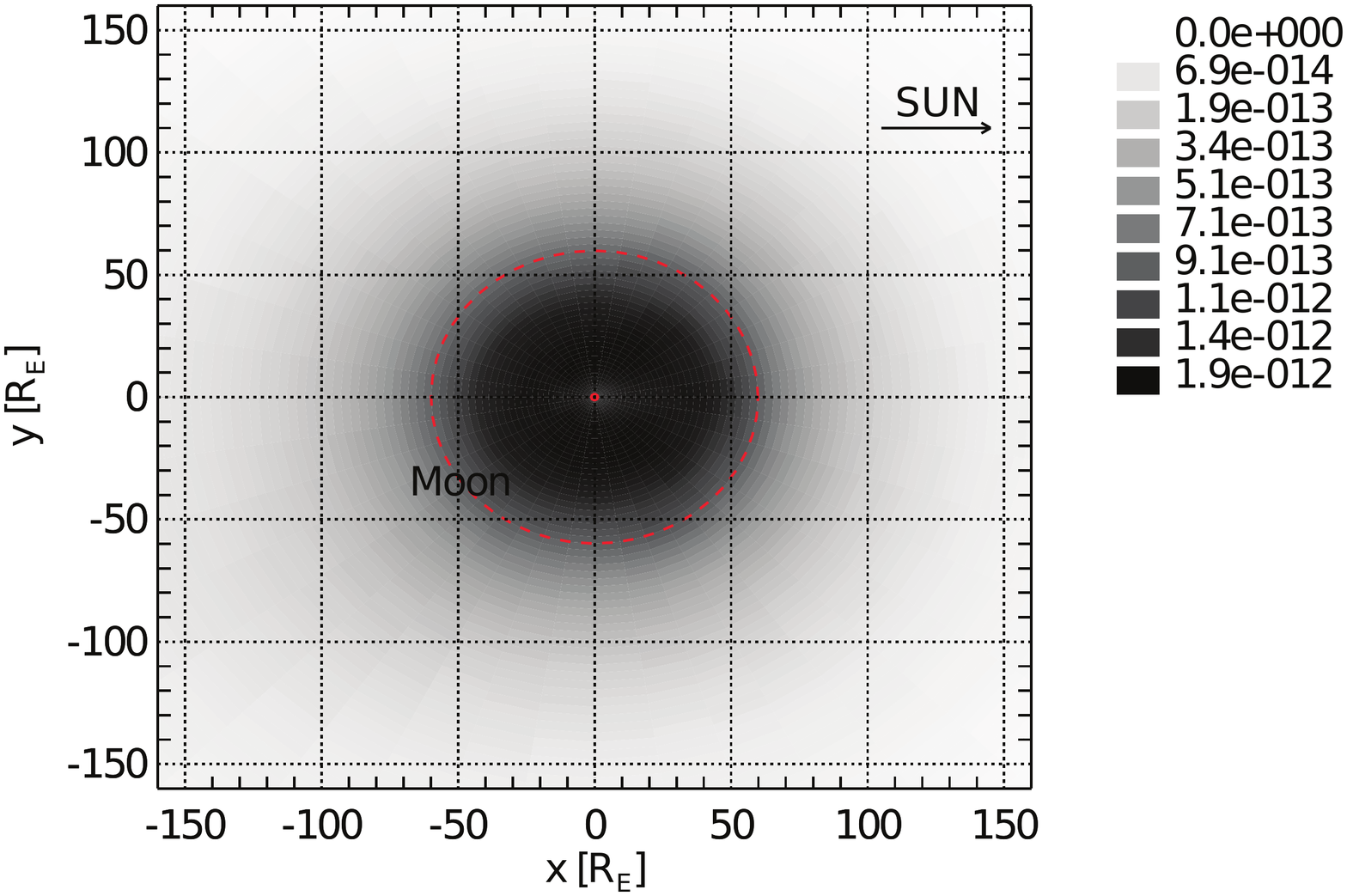}
    \includegraphics[width=0.9\columnwidth]{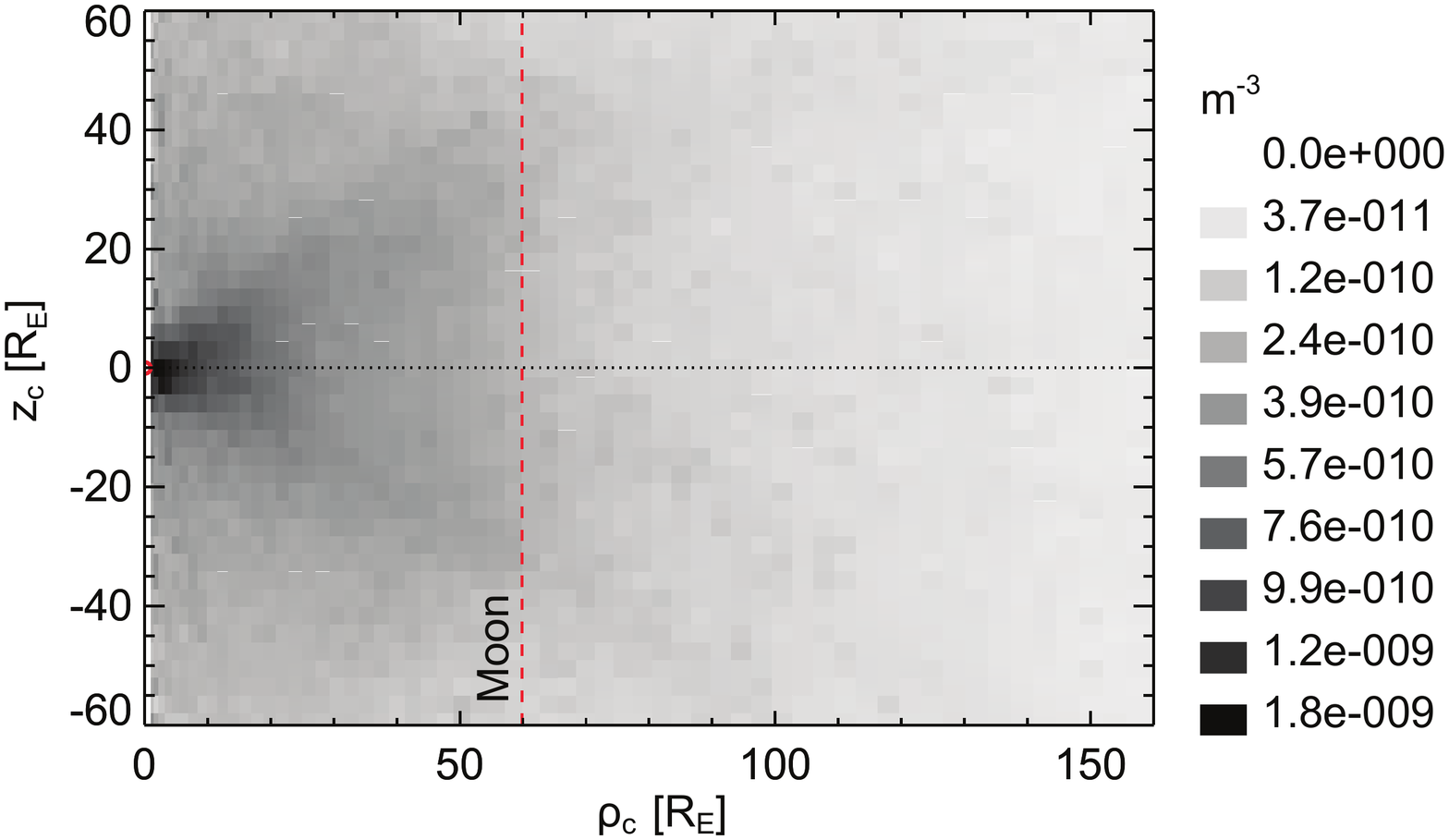}
    \caption{Optical depth and number density of particles ejected from the surface of the Moon, obtained from numerical simulations. The red circle denotes the Earth, and the red dotted line denotes the orbit of the Moon. \emph{Upper Panel}: Normal geometric optical depth projected in the Earth's equatorial plane (the $x-y$ plane). Here the positive $x$ axis points in the direction of the Sun. \emph{Lower Panel}: Azimuth-averaged number density in the $\rho_c-z_c$ plane, where $\rho_\mathrm{c}=\sqrt{x^2+y^2}$, $z_\mathrm{c}=z$, and the $z$ axis is the normal of the $x-y$ plane. Reproduced with permission from \cite{yang2021}, \copyright ESO.}
    \label{fig:opticaldepth}
\end{figure}

 \cite{yang2021} considered the gravitational forces from the Sun, the Earth including its $J_2$ term, and the Moon, and the solar radiation pressure in the lunar dust dynamics modelling, and obtained the average lifespans for grains with different sizes. The solar radiation pressure and the Lorentz force have stronger effects on particles smaller than $<10\,\mathrm{\mu m}$ due to their scaling with the particle size, and thus their trajectories are more likely to become hyperbolic orbits. As shown in Fig.~\ref{fig:LifeSpan}, small particles escape quickly with short lifespans (usually several weeks) in the Earth-Moon system, while larger dust particles can move relatively stably in the Earth-Moon system and thus have longer average lifespans \citep[up to one year;][]{yang2021}. The authors also derived the steady-state configuration of particles in the Earth-Moon system that ejected from the lunar surface in hypervelocity impacts of micrometeoroids, over an evolution time of more than $100$ years. Those particles occupy a ring between the Earth and the Moon, which is slightly offset toward the Sun (see Fig.~\ref{fig:opticaldepth}).

\section{Summary and discussion}

This paper mainly reviews the research on the lunar dust dynamics as follows,

1) The interplanetary impactors for the Earth-Moon system consist of six interplanetary meteoroid sources (HE, AH, AP, AA, NT and ST), which are mainly generated from the JFCs, HTCs and OCCs, as well as their disruptions. Among them, the AP sources are the dominant producer, which determines the asymmetry of the observed lunar dust cloud. The short-period comets (JFCs) produce 1.3 times more meteoroids than long-period comets (like HTCs and OCCs) while the meteoroids from long-period comets generate more ejecta dust particles due to their higher impacting speeds. 

2) Detection missions for lunar dust have been carried out in the past few decades since Apollo era. The LDEX onboard the LADEE firstly reported an asymmetric and permanent dust cloud around the Moon. Chang'e-3 recorded the amount of splashed dust caused by the landing process and the total accumulation mass, which is comparable to that from Apollo missions. Over the next decade, several lunar exploration missions are planned to be launched, to detect the lunar surface and near-lunar space environment, such as the Artemis, the Luna 25-27, the Chandrayaan 3, the Tanpopo experiment, and the EQUULEUS.

3) The mass production rate for particles ejected from the lunar surface is dependent on the relative strength of each source and the ejecta position of particles on the surface. The number of particles from the lunar surface for each size follows a power-law distribution with the exponent $0.9$, obtained from the LDEX in-situ measurement. The initial speed also follows a power law and the plausible exponent for particles from the lunar surface is $2.2$.

4) Particles generated from the lunar surface are mainly influenced by the gravity of Earth, Moon and Sun, the Poynting-Robertson drag, the solar radiation pressure and the Lorentz force. Among them, nongravitational forces, including the solar radiation pressure and the Lorentz force, have more significant effects on small particles. Therefore, small particles escape the Earth-Moon system quickly with weeks of life while larger particles have longer average lifetimes of up to one year. Most of lunar dust particles fall back to the lunar surface while some can escape the lunar gravity and evolve into the Earth-Moon system. After long-term evolution, these particles occupy a torus between the Earth and the Moon.

However, there are still some problems and issues related to lunar dust dynamics that need to be solved. To the knowledge of the authors, at the time of writing this review article, some effects that may influence the dynamics of particles in the Earth-Moon system are not yet taken into consideration: (1) The higher degree and order spherical harmonics of the Earth's gravity field. (2) The temporal variation of the magnetic field which is important for small particles. The IMF and the Earth's magnetic field are dynamic, i.e. both vary with the time \citep{mikic1999magnetohydrodynamic, russell1993magnetic}, which introduces difficulties for the dust dynamics modeling. (3) The charging process of particles in the plasma environment that will affect the Lorentz force, which is also important for small particles (especially for submicron and nano particles). (4) The sputtering of particles by the ambient environment, which reduces the grain size and then affect the dust dynamics. The sputtering is important only for the particles with long lifetimes \citep{hu2019thermal}. (5) Resonance. Particles may be trapped into resonant positions in the Earth-Moon system due to the drag forces \citep{hamilton1994comparison}. These effects may be important for the dust dynamics in the certain size range or in the certain space regions of the Earth-Moon system.
 
 For the levitation process of particles from the lunar surface, the electrostatic force is generally believed to be the main reason for the separation \citep[e.g.,][]{popel2018lunar,yeo2021dynamics}, but the Van der Waals adhesive force is rarely considered. The adhesive force for large particles ($>10^3\,\mathrm{\mu m}$) is negligible, but it dominants for micron and submicron particles when separating from the lunar surface \citep{hartzell2011role}. Unfortunately, the value of adhesive force is hard to estimate due to the great differences in shape and size of particles, and the difficulty in characterizing the surface roughness \citep{dove2011mitigation,zakharov2021lunar}. Besides, the thermodynamics is not taken into consideration, which may affect the lofting process of lunar particles. For extremely small particles (nanoparticles), the thermal energy may be larger than the adhesive energy, i.e., the high temperature weakens the links between different particles. In this case, the “boiling” particles are levitated from the lunar surface \citep{rosenfeld2016lunar}. Moreover, the landing process of the lunar landers causes a large number of splashed particles \citep{zhang2020situ}, and the instability of the natural electrostatic fields \citep{zakharov2021lunar}. The influence of human exploration activities on the lunar dust dynamics needs further research.
\begin{ethics}
\begin{conflict}
There are no conflicts of interest to declare.
\end{conflict}
\end{ethics}

\begin{authorcontribution}
All authors contributed significantly to this work. The first draft of the manuscript was written by Kun Yang and all authors commented on previous versions of the manuscript.
\end{authorcontribution}

\begin{dataavailability}
The data are available from the corresponding author on reasonable request.
\end{dataavailability}

\begin{fundinginformation}
This work was supported by National Natural Science Foundation of China (No. 12002397) and by the China Scholarship Council (CSC, 201906220134). 
\end{fundinginformation}

\bibliographystyle{spr-mp-nameyear-cnd}
\bibliography{biblio-u1}

\end{document}